\documentclass[prl,aps,twocolumn,groupedaddress,floatfix,showpacs]{revtex4-1}
\usepackage{bm}

\usepackage{graphicx,amssymb,latexsym,amsmath,amsbsy,bm}

\newcommand{\beq}{\begin{equation}}
\newcommand{\eeq}{\end{equation}}
\newcommand{\bea}{\begin{eqnarray}}
\newcommand{\eea}{\end{eqnarray}}

\newcommand{\nn}{\nonumber}
\newcommand{\benn}{\begin{displaymath}}
\newcommand{\eenn}{\end{displaymath}}
\newcommand{\DFT}{\textsc{dft}}

\newcommand{\TDSLDA}{\textsc{tdslda}}
\newcommand{\BEC}{\textsc{bec}}

\newcommand{\BCS}{\textsc{bcs}}
\newcommand{\BdG}{\textsc{b}d\textsc{g}}

\newcommand{\SLDA}{\textsc{slda}}

\newcommand{\QMC}{\textsc{qmc}}

\newcommand{\UFG}{\textsc{ufg}}

\newcommand{\GP}{\textsc{gp}}
 
\newcommand{\QSW}{\textsc{qsw}} 
\newcommand{\DW}{\textsc{dw}}

\begin{document}

\title{\large  Quantum Shock Waves and Domain Walls in the Real-Time Dynamics of a Superfluid Unitary Fermi Gas }

\author{ Aurel Bulgac$^1$, Yuan-Lung (Alan) Luo$^1$, and Kenneth J. Roche$^{2,1}$  }
\affiliation{$^1$Department of Physics, University of Washington, Seattle, WA 98195--1560, USA} 
\affiliation{$^2$Pacific Northwest National Laboratory, Richland, WA 99352, USA}

\begin{abstract} 
We show that in the collision of two superfluid fermionic atomic clouds one observes the formation of quantum shock waves  as discontinuities in the number density and collective flow velocity. Domain walls, which are topological excitations of the superfluid order parameter, are also generated and exhibit abrupt phase changes by $\pi$ and slower motion than the shock waves. The domain walls are distinct from the gray soliton train or number density ripples formed in the wake of the shock waves and observed in the collisions of superfluid bosonic atomic clouds. Domain walls with opposite phase jumps appear to collide elastically. 
\end{abstract}

\date{\today}       


% 03.75.Ss Degenerate Fermi gases
% 03.75.Lm Tunneling, Josephson effect, Bose-Einstein condensates in periodic potentials, solitons, vortices, and topological excitations
% 03.75.-b 	Matter waves (for atom interferometry, see 37.25.+k; see also 67.85.-d ultracold gases, trapped gases in quantum fluids and solids)
%  03.75.Kk 	Dynamic properties of condensates; collective and hydrodynamic excitations, superfluid flow  

\maketitle

%%%%%%%%%%%%%%%%%%%%%%%%%%%%%%%%%%%%%%%%%%%%%%%
  
Quantum shock waves (\QSW s) and solitons have been observed in a dilute atomic Bose gas a decade ago  \cite{Dutton:2001}, a result which generated a flurry of experimental and theoretical work \cite{Simula:2005,Hoefer:2006,Chang:2008,Carretero-Gonzalez:2008}. In classical hydrodynamics, shock waves appear as discontinuities in density, flow velocity, temperature and other characteristics of the fluid.  In a fluid in motion one can observe another remarkable structure, solitons. Solitons are usually a manifestation of the competition between the dispersive mechanisms and nonlinearities at work in a fluid. While dispersion effects tend to smooth out discontinuities, nonlinearities sometimes oppose this effect instead, stabilizing a large local density variation \cite{Whitham:1974} that leads to the appearance of a bright, dark or a gray soliton.  The quantum fluid dynamics of a  superfluid dilute atomic Bose gas is typically described using the nonlinear Gross-Pitaevskii (\GP) equation, and at sufficiently low temperatures there does not appear to be any need for dissipative processes. The accurate treatment of the real-time dynamics of a superfluid Unitary Fermi Gas (\UFG) requires a more complex approach, using an extension of the Density Functional Theory (\DFT) to superfluid systems and time-dependent phenomena, see Refs.  \cite{BY:2003,Bulgac:2005a,Bulgac:2009,Bulgac:2011,Bulgac:2011b}.

\begin{figure}[ht]
\includegraphics[width=0.43\textwidth]{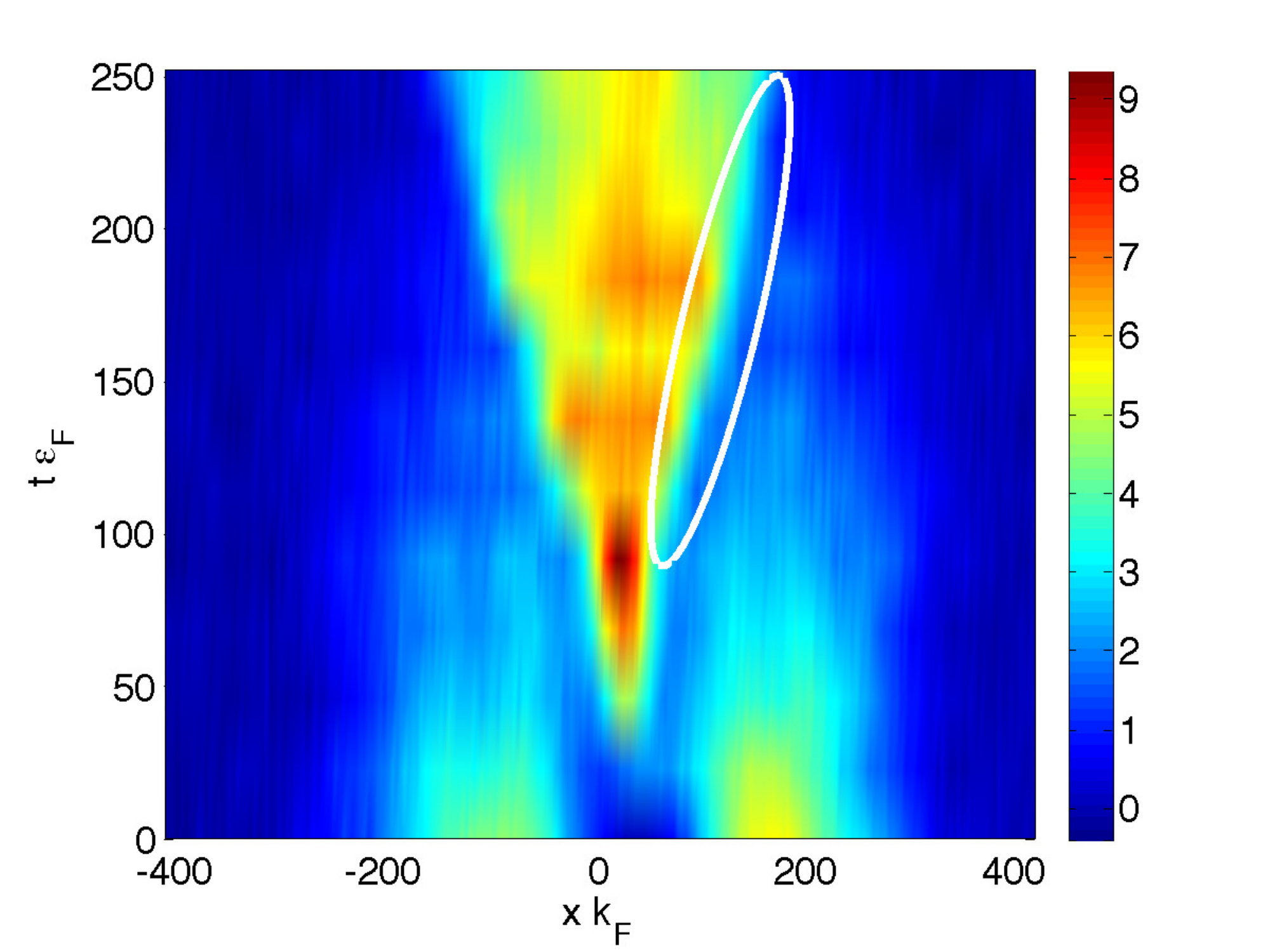}
\caption{ (Color online) The space-time evolution of the 1D density profiles along the trap axis as reported in Fig. 2 of Ref. \cite{Joseph:2011}. We have outlined part of the shock wave front with a white oval here and in Figs. \ref{fig:rho}, \ref{fig:delta}, \ref{fig:phase}, and \ref{fig:vx}.  }
\label{fig:exp}  
\end{figure}

Recently, a new experiment reported on the observation of \QSW s  in a  \UFG\ \cite{Joseph:2011}, see Fig. \ref{fig:exp}. The \UFG\  is a system exactly in the middle of the \BCS-\BEC\ crossover and as such its properties often qualitatively interpolate between those of a Fermi and a Bose superfluid. However, while BCS and BEC systems are weakly interacting, the \UFG\ is a strongly interacting superfluid in which the critical temperature and the Landau critical velocity attain their highest values (in appropriate units).  The velocity equation used in Ref. \cite{Joseph:2011} to model shock waves in the collisions of two \UFG\ clouds neglects the ``quantum pressure term" and a phenomenological viscosity term is added:
%%%%%%%%%%%%%%%%%%%%%%%%%%%%%%%%%%%%%%%%%%%%%%% 
%\beq
%\dot{n} + {\bm \nabla}\cdot ({\bf v}n)=0, \nn 
%\eeq
\beq
n\dot v_k+n\nabla_k\left \{ \frac{{\bf v}^2}{2} 
+\mu[n]+V_{ext}\right \} +\nu \nabla_l \{n  [\nabla \odot v  ]_{kl}\}
=0,\nn
\eeq
%%%%%%%%%%%%%%%%%%%%%%%%%%%%%%%%%%%%%%%%%%%%%%% 
in which one can choose the mass of the atom $m=1$, $v_k$ and $\nabla_k$ are the cartesian coordinates of ${\bf v}$ and ${\bm \nabla}$ respectively, and $[\nabla \odot v]_{kl} = \nabla_k v_l+\nabla_l v_k-2/3\delta_{kl}{\bm \nabla}\cdot{\bf v}$.  Above  $\mu[n]$ is the chemical potential in homogeneous matter at a given number density $n$, and we have suppressed the explicit dependence of the number density $n({\bf r},t)$, velocity ${\bf v}({\bf r},t)$ and external trapping potential $V_{ext}({\bf r},t)$ on space-time coordinates.  In the superfluid phase both the shear viscosity and the ``quantum pressure'' terms are proportional to $\hbar$ and a priori it is not obvious that one can neglect one term, but not the other. Unlike the case of dilute Bose gases, where the \QSW s were interpreted as ``pure'' dispersive shock waves \cite{Hoefer:2006} with no need for dissipative effects, the results of the experiment \cite{Joseph:2011} on \UFG\ received an interpretation similar to classical shock waves, in which dissipation plays a crucial role in the formation of the shock wave front. 
%-------------------------------------------------------------
%-------------------------------------------------------------
%-------------------------------------------------------------
%-------------------------------------------------------------
These two distinct interpretations of the experiments on Bose and Fermi dilute gases are difficult to reconcile. In the \BEC\ regime, the role of dissipation is negligible (at least at the phenomenological level) and the \QSW\ and the density ripples identified with soliton trains can be described by dispersive effects alone. Viscosity was introduced phenomenologically in Ref. \cite{Joseph:2011}, to a large extent in order to avert the onset of a ``gradient catastrophe''  \cite{Bettelheim:2006}. At the same time, the \UFG\ is widely accepted as a prime example of an almost perfect fluid \cite{Cao:2011a}. In the \UFG\, the bulk viscosity vanishes and the shear viscosity is at a minimum as a function of the coupling constant across the BEC-BCS crossover, see Refs. \cite{Cao:2011a,Enss:2011,Gabriel:2012} and references therein,  and thus we see no compelling theoretical arguments to include it in the present study. In this respect, the present approach is similar to other studies of dilute Fermi gases \cite{Antezza:2007a,Scott:2011,Spuntarelli:2011,Liao:2011,Salasnich:2011}.  Another significant limitation of the hydrodynamic approach, which was used in Ref. \cite{Joseph:2011},  is its inability to describe quantum topological excitations (quantized vortices, domain walls (\DW)), both of which have been observed in the similar experiments with bosons \cite{Dutton:2001,Simula:2005,Hoefer:2006,Chang:2008,Carretero-Gonzalez:2008}.

In the case of colliding \UFG\ clouds, we observe the generation of both \QSW s and \DW s. The \DW\ excitation has been suggested in other  simulations \cite{Antezza:2007a,Scott:2011,Spuntarelli:2011,Liao:2011}. The \DW s are excitations of the superfluid order parameter and not the number density ripples identified as soliton trains trailing the wake of the shock waves, as discussed in Refs. \cite{Dutton:2001,Simula:2005,Hoefer:2006,Chang:2008,Carretero-Gonzalez:2008}. We will make this distinction in order to avoid confusion. We show that the number density of two colliding \UFG\ clouds shows a behavior very similar to the one observed in experiment  \cite{Joseph:2011}.  In the wake of the \QSW s we observe in addition the formation of \DW s.  The \DW s emerge as quite sharp changes in the phase of the superfluid order parameter by $\pi$, and are correlated with minima of the number density.  \DW s propagate through the system at slower speeds than the \QSW s and are topological excitations  similar to quantum vortices. \DW s always appear in pairs with opposite jumps of the order parameter phase and appear to collide essentially elastically with one another and with the system boundary. These phenomena are observed in the absence of any dissipation, which is expected to play a negligible role at temperatures close to absolute zero, see Ref.  \cite{Enss:2011,Gabriel:2012} and references therein.

The extension of the \DFT\ to superfluid fermionic systems and time-dependent phenomena has been described and applied to a number of phenomena in nuclear physics and the physics of cold gases \cite{Bulgac:2009,Bulgac:2011,Bulgac:2011b}. This approach is known as the Superfluid Local Density Approximation (\SLDA) and \TDSLDA\ for its time-dependent version respectively, based on the simplest possible (un-renormalized) energy density functional:
%%%%%%%%%%%%%%%%%%%%%%%%%%%%%%%%%%%%%%%%%%%%%%% 
\beq
\mathcal{E}=\frac{\hbar^2}{m}\left [
\alpha \frac{\tau}{2} + \beta \frac{3 (3 \pi^2)^{2/3} n^{5/3}}{10} + \gamma\frac{ |\nu |^2}{n^{1/3}} \right ] +V_{ext}n
, \nn
\eeq
%%%%%%%%%%%%%%%%%%%%%%%%%%%%%%%%%%%%%%%%%%%%%%% 
where $\tau_c = 2 \sum_{E_n < E_c} |{\bf \nabla} v_n|^2 $, $n = 2 \sum_{E_n < E_c} |v_n|^2$, and $\nu_c = \sum_{E_n < E_c} v_n^{\ast} u_n$ are the kinetic, number and anomalous densities respectively,  and $V_{ext}$ is an external one-body potential. The dynamical evolution of the system is described by Bogoliubov-de Gennes-like equations (\BdG) for the quasiparticle wave functions (qpwfs) $(u_n,v_n)$
%%%%%%%%%%%%%%%%%%%%%%%%%%%%%%%%%%%%%%%%%%%%%%% 
\beq
i \hbar \frac{\partial}{\partial t}
\left( \begin{array}{c} u_n \\ v_n \end{array} \right)
=
\left( \begin{array}{cc} h  & \Delta \\ \Delta^{\ast} & -h  \end{array} \right)
\left( \begin{array}{c} u_n \\ v_n \end{array} \right), \nn
\eeq
%%%%%%%%%%%%%%%%%%%%%%%%%%%%%%%%%%%%%%%%%%%%%%% 
where the single-particle Hamiltonian, $h$, and pairing potentials, $\Delta $, are obtained by taking the appropriate functional derivatives of the energy  density functional $\mathcal{E}$.  The dimensionless constants $\alpha$, $\beta$ and $\gamma$ are fixed by the energy per particle, pairing gap and quasiparticle spectrum obtained from Quantum Monte Carlo (\QMC) calculations of the homogeneous infinite system.  For the description of the renormalization procedure, various technical details, and numerical implementation details see Refs. \cite{Bulgac:2011,Bulgac:2011b}. Within \SLDA\ various properties of a \UFG\ are reproduced within a few percent accuracy, mainly limited so far by the accuracy of the \QMC\ values for the pairing gap and for the effective mass \cite{Bulgac:2011}. Dimensional and symmetry arguments,  renormalizability, and Galilean invariance define \TDSLDA\ uniquely and with an overall accuracy not worse than 10\%.  Unlike the \BdG\ approximation, in which interaction effects vanish in the absence of pairing correlations, the energy of both superfluid and normal phases at zero temperature are described accurately. The \UFG\ has a condensation energy  $\approx  20\%$ of the total interaction energy. In the \BdG\ approximation, however, the condensation energy is equal to the interaction energy in the case of an {\UFG}. Even at unitarity, the meanfield energy dominates the pairing energy, and that affects the dynamics accordingly. An \UFG\ initially in a superfluid phase that is subjected to an external time-dependent agent can become normal and the Cooper pairs could be destroyed, but the particles will still be strongly interacting. In contradistinction, in the \BdG\ approximation, which was used in studies of solitons \cite{Antezza:2007a,Scott:2011,Spuntarelli:2011,Liao:2011}, the normal phase is simulated as a non-interacting Fermi gas \cite{Bulgac:2011b}.

%-------------------------------------------------------------
%-------------------------------------------------------------
\begin{figure}[ht]
\includegraphics[width=0.43\textwidth]{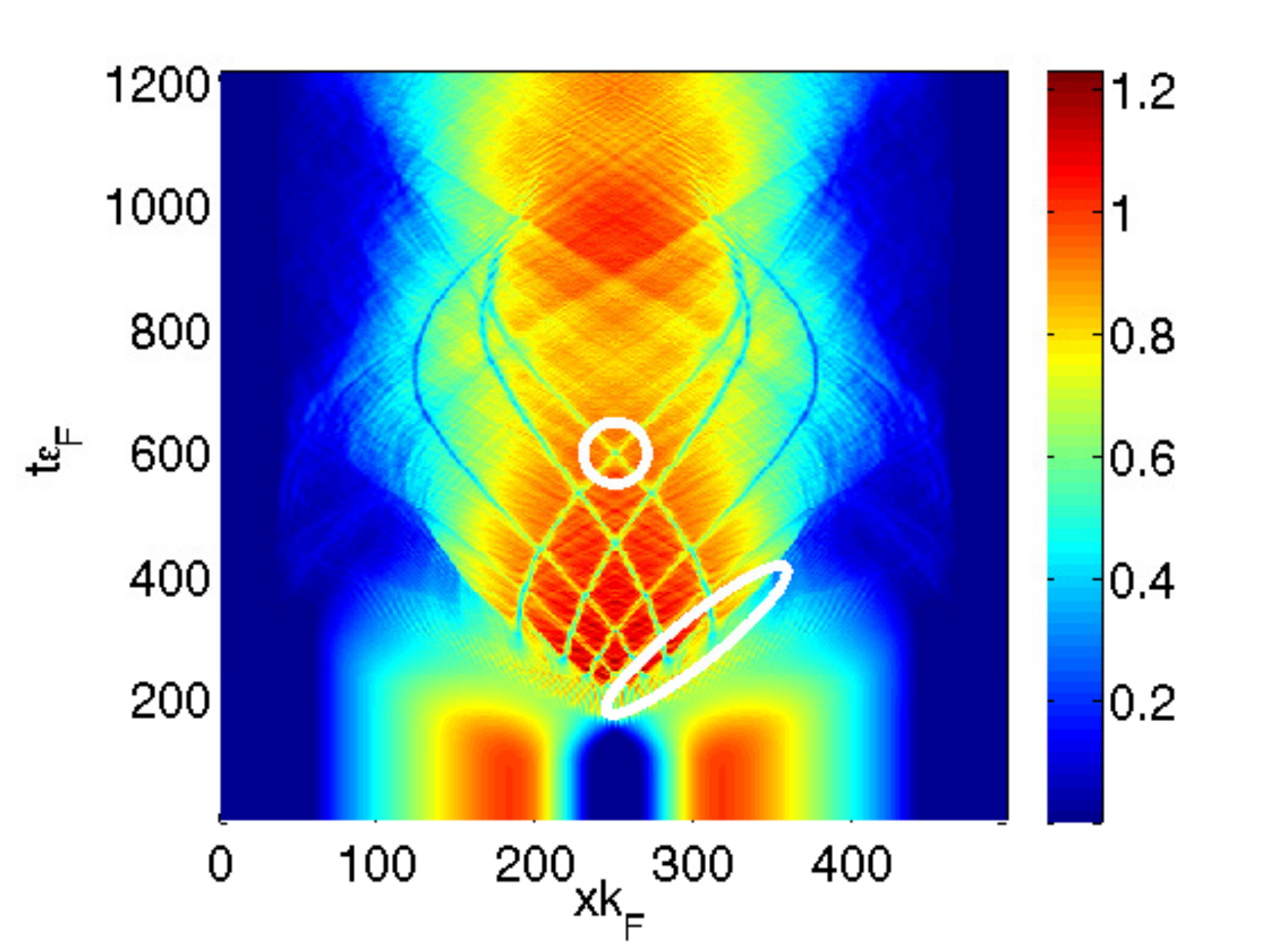}
\caption{  (Color online) The space-time evolution of the number density profile $n(x,0,t)$ along the collision axis in TDSLDA simulation of the collision of two \UFG\ clouds.  Here $k_F$ and $\varepsilon_F=k_F^2/2$ ($\hbar=m=1$) are the initial values of the Fermi wave vector and energy at the center of clouds. The region where one elastic collision of two \DW s occurs is outlined with a white circle here and in Figs. \ref{fig:delta}, \ref{fig:phase}, and \ref{fig:vx}.  }
\label{fig:rho}  
\end{figure}
%-------------------------------------------------------------
%-------------------------------------------------------------

\begin{figure}[ht]
\includegraphics[width=0.15\textwidth]{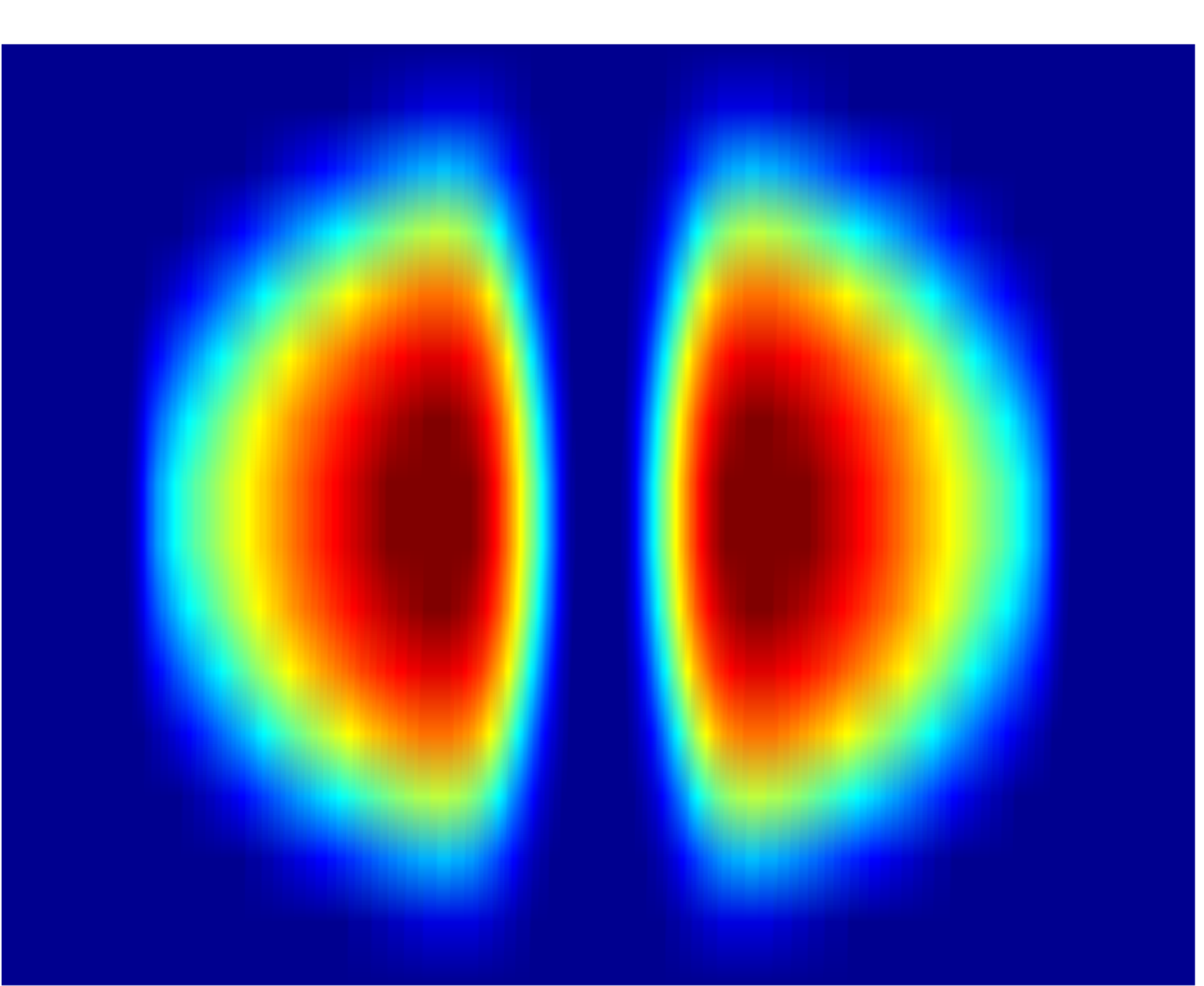}
\includegraphics[width=0.15\textwidth]{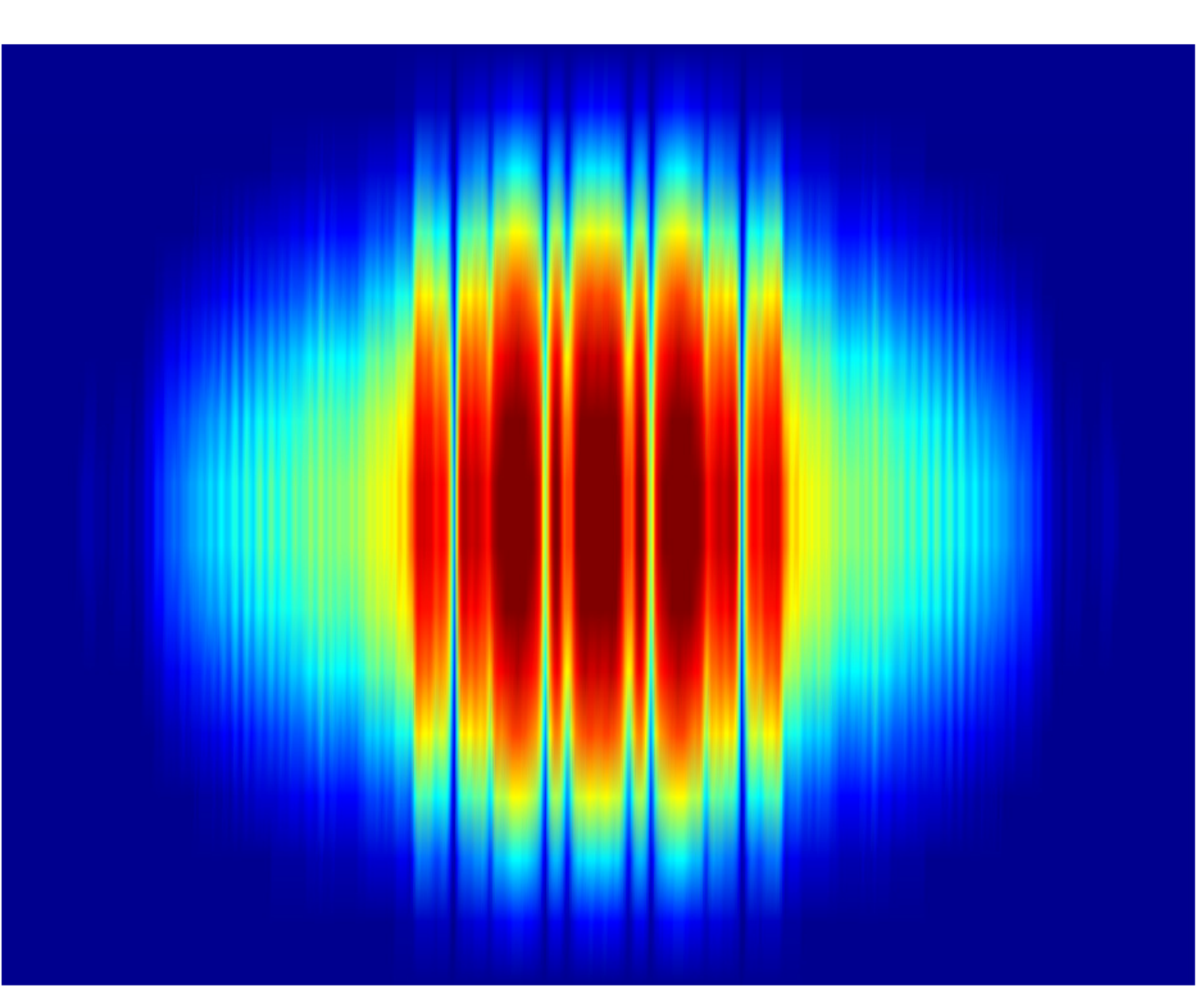}
\includegraphics[width=0.15\textwidth]{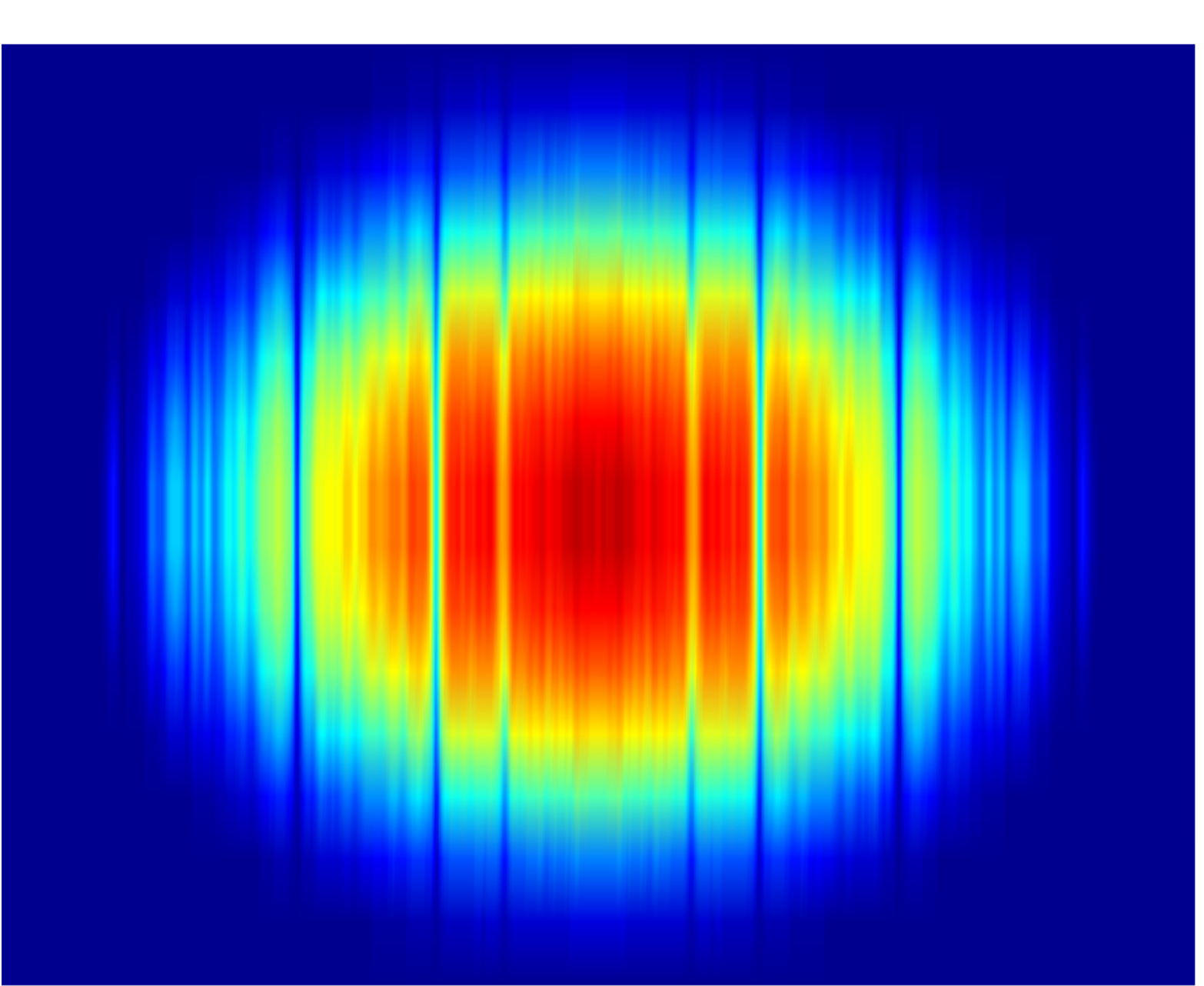}
\caption{ (Color online) Three consecutive frames showing the absolute magnitude of the pairing field $|\Delta(x,y,t)|$  in the $xy$-plane at times $t\varepsilon_F=30, \; 350,\; 690$, see also  Figs. \ref{fig:rho} and \ref{fig:delta}. The $x$- and $y$-directions (not shown to scale here) have an aspect ratio of $\approx 30$. By the time of the second frame the \QSW s have emerged from the cloud. \DW s appear as planar number density depletions with a width comparable to the diameter of a quantum vortex, and can be highly visible using standard techniques \cite{ZA-SSSK:2005lr}.  We have observed the formation of \DW s so far only in traps with elongations larger than in experiment \cite{Joseph:2011}. In simulations of clouds with smaller aspect ratios we observe only \QSW s with density profiles very similar to Ref. \cite{Joseph:2011}. }
\label{fig:frames}  
\end{figure}

\begin{figure}[ht]
\includegraphics[width=0.43\textwidth]{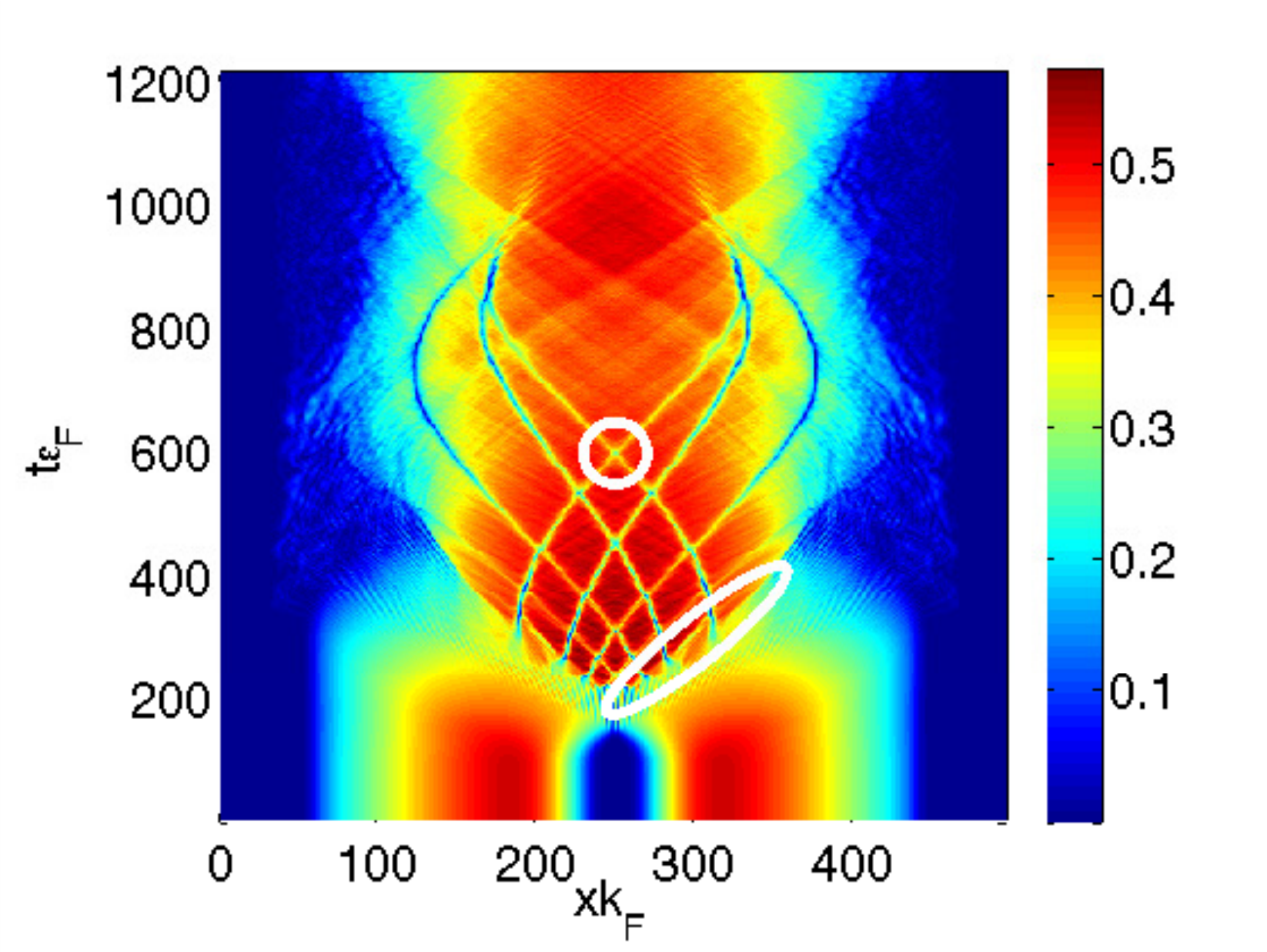}
\caption{  (Color online) The  magnitude of the pairing gap $|\Delta(x,0,t)|$ corresponding to Fig. \ref{fig:rho}. The \DW s appear as significant depletions of the pairing field, and always appear in pairs with opposite phase jumps, see also Figs. \ref{fig:phase} and \ref{fig:vx}.   }
\label{fig:delta}  
\end{figure}

\begin{figure}[ht]
\includegraphics[width=0.43\textwidth]{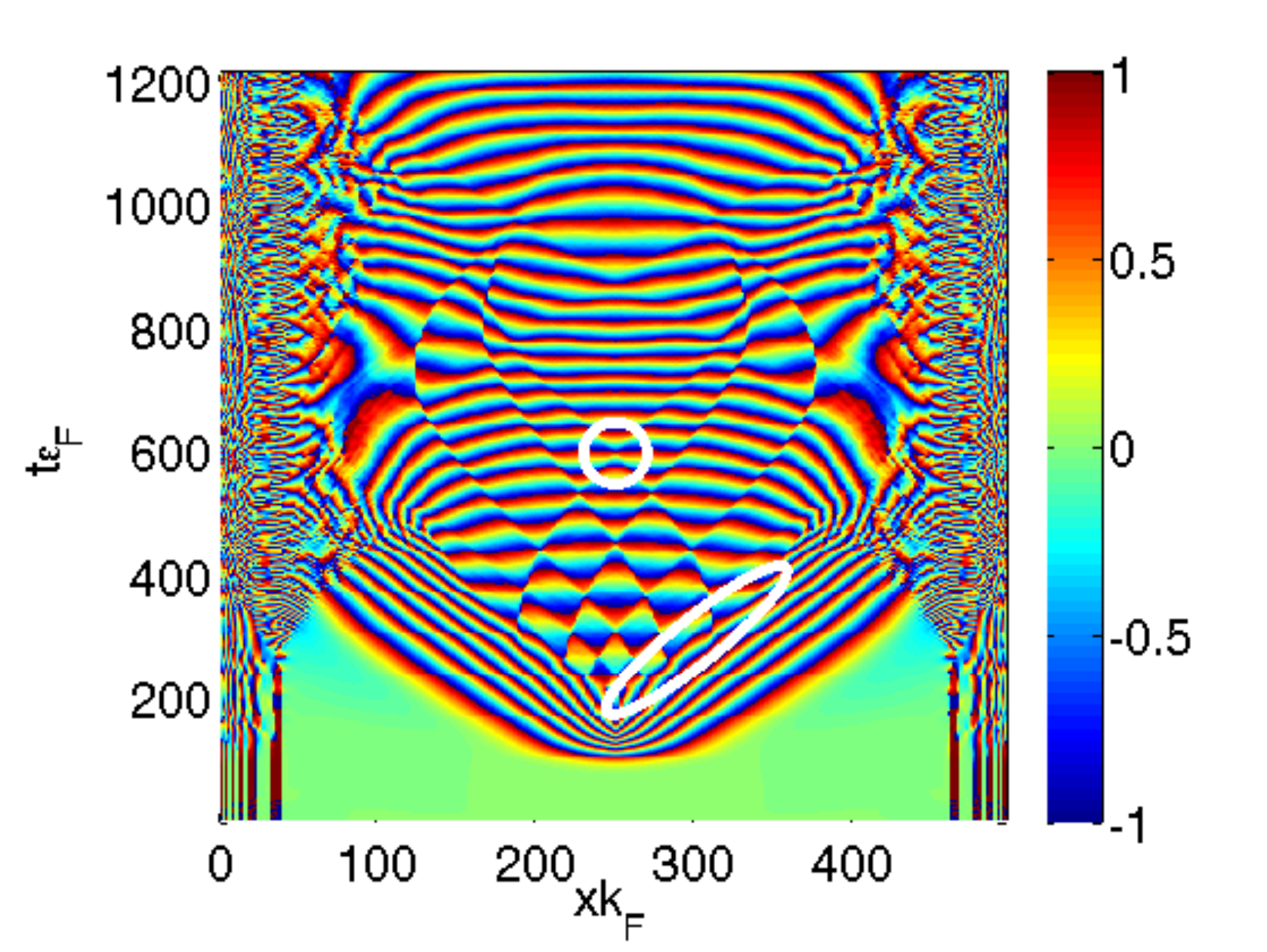}
\caption{  (Color online) The phase of the pairing gap $\arg \Delta(x,0,t)/\pi$.  }
\label{fig:phase}  
\end{figure}

\begin{figure}[ht]
\includegraphics[width=0.43\textwidth]{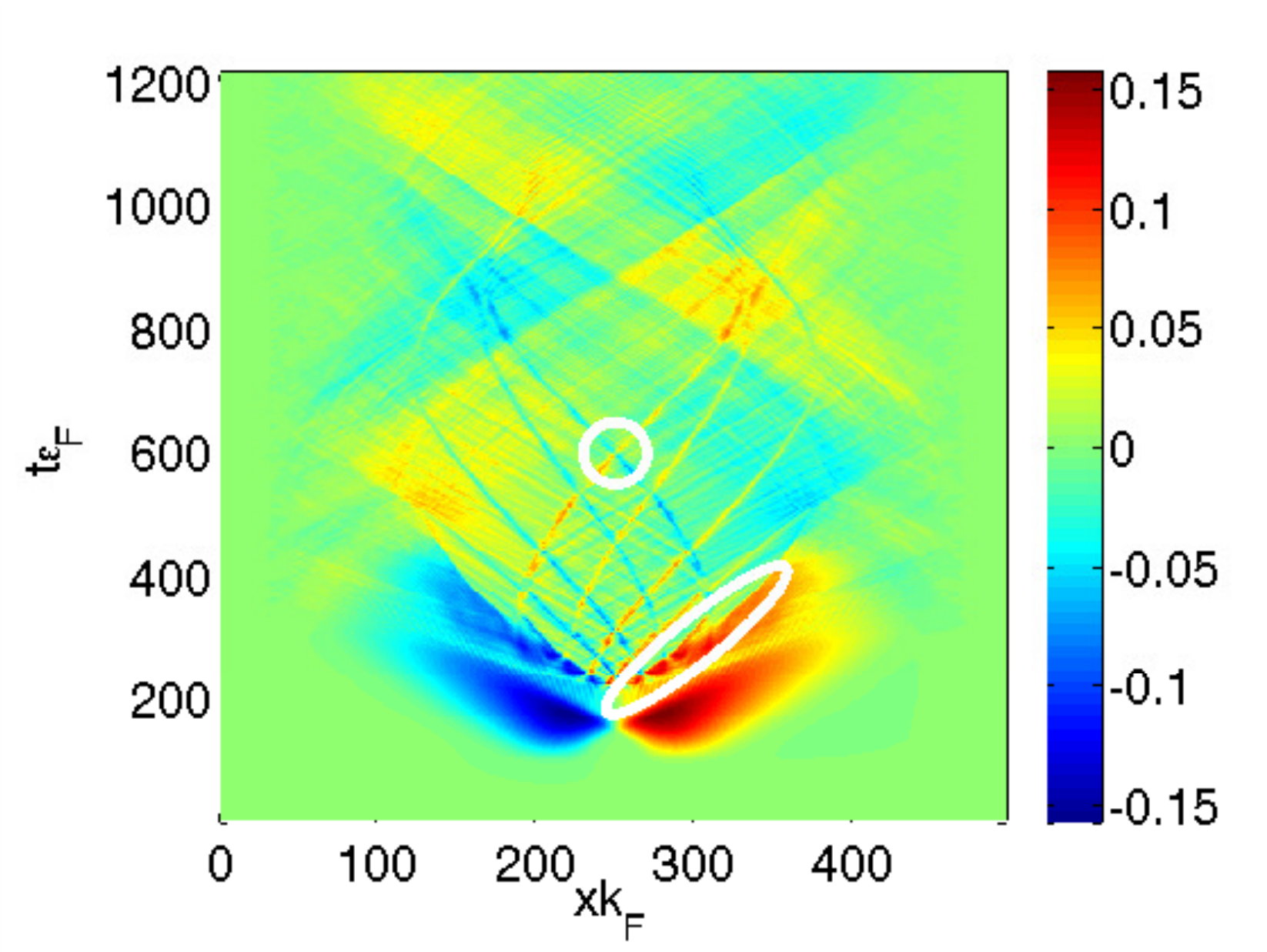}
\caption{ (Color online) The $x$-component of the collective flow velocity field $v_x(x,0,t)$ along the axis of collision. At the front of the two shock waves the velocity field  undergoes a rapid change, and the matter flows in opposite directions.   }
\label{fig:vx}  
\end{figure}

We have performed simulations of the cold atom cloud collisions assuming that qpwfs have the structure  $u_n({\bf r},t)\Rightarrow\exp(ik_{nz}z)u_n(x,y,t),v_n({\bf r},t)\Rightarrow\exp(ik_{nz}z)v_n(x,y,t)$ with periodic boundary conditions in the $z$-direction and a rather stiff  harmonic confining potential in the $y$-direction.  The time-dependent trapping potential along the collision $x$-axis had a similar profile and time dependence as in experiment \cite{Joseph:2011}. The solitons and the shock waves now are 2D in character, their stability properties are slightly different than in 3D,  and the sound velocity is modified \cite{Capuzzi:2006}. Typical results of these simulations are shown in Figs. \ref{fig:rho}, \ref{fig:frames}, \ref{fig:delta}, \ref{fig:phase}, and \ref{fig:vx}. One can zoom into the online figures in order to see details, which otherwise would escape the naked eye.  
%-------------------------------------------------------------
%-------------------------------------------------------------
%-------------------------------------------------------------
%-------------------------------------------------------------

%-------------------------------------------------------------
%-------------------------------------------------------------
%-------------------------------------------------------------
%-------------------------------------------------------------

The simulation results, Fig. \ref{fig:rho}, show remarkable similarities with the experiment, Fig. \ref{fig:exp}. The \QSW s speed in experiment ($v_{qsw}/v_F\approx 0.35$) and simulations ($v_{qsw}/v_F\approx 0.25$) agree within $\approx 25-30\%$. The differences can be ascribed to various experimental uncertainties (in particular the particle number), different trap shapes ($V_{trap}(x,y,z) \propto \omega^2_\parallel x^2+\omega^2_\perp(y^2+z^2)$ and $V_{trap}(x,y,z)\propto \omega^2_\parallel x^2+ \omega^2_\perp y^2$ in experiment and simulations respectively) and sizes, different cloud aspect ratios, and also different amount of initial collisional kinetic energy injected into the system. In spite of being confined in the $y$-direction in a harmonic potential, the \DW\ are planes perpendicular to the collision $x$-axis, see Fig. \ref{fig:frames}. However, in Ref. \cite{Joseph:2011} the experimental set-up prevented the authors from observing the \DW s. The images corresponding to various frames reported there were taken in different realizations of the two colliding clouds. On one hand, the phase differences of the two initially separated condensates are likely random and maybe difficult to control, see Ref. \cite{Andrews:1997}. On the other hand, the density profile fluctuations from shot-to-shot  in Ref. \cite{Joseph:2011} point to a rather low spatial resolution attained in these measurements (see Fig. 2 in Ref.  \cite{Joseph:2011}), which explains why \DW s have not been observed in this experiment.  We have performed simulations by varying the initial relative phase of the condensates. While the overall picture of the collisions remains unchanged, the number of \DW s created varies. The density ripples in the wake of the shock waves discussed in experiments with Bose dilute clouds  \cite{Dutton:2001,Simula:2005,Hoefer:2006,Chang:2008,Carretero-Gonzalez:2008} and interpreted there as a soliton train, are formed here as well. By zooming in the online Figs. \ref{fig:rho} and  \ref{fig:delta} one can notice that before the shock wave is formed well defined matter wave interference occurs. The discontinuity in the number density and order parameter at the wake of the shock wave is accompanied by a similar discontinuity in the collective flow velocity, see Fig. \ref{fig:vx}.  The \DW s which form in the wake of the shock wave have lower speeds. The width of the domain wall density depletion is comparable with the diameter of the density depletion of the quantized vortex core in a UFG \cite{BY:2003,Bulgac:2005a}. This aspect might present a challenge if one were to attempt a direct observation. However, the experimental technique implemented to visualize the quantum vortex lattice in a UFG \cite{ZA-SSSK:2005lr} can be implemented to put in evidence \DW s, particularly since a spatial imprint left by a single \DW s is a much larger 2D-structure, as compared to the 1D imprint  left by vortex core,  see Fig. \ref{fig:frames}.  In sufficiently wide traps \DW s might develop ``snake'' instabilities as in Ref. \cite{Dutton:2001}, which would be a clear ``smoking gun'' of \DW s formation.

%-------------------------------------------------------------
%-------------------------------------------------------------
%-------------------------------------------------------------
%-------------------------------------------------------------

%-------------------------------------------------------------
%-------------------------------------------------------------
%-------------------------------------------------------------
%-------------------------------------------------------------

In summary, we have shown that \QSW s and \DW s are formed in the collision of two initially independent superfluid fermionic clouds. The \QSW s manifest themselves as rather sharp discontinuities in the number density, superfluid order parameter and collective flow velocity.  They reflect essentially elastically from system boundaries, but with significant reduction in their intensity. The shock waves lead to the formation of \DW s, topological excitations of the superfluid order parameter similar to quantum vortices. As in the case of vortices, \DW s show-up as significant depletions of the number density, but over much wider spatial regions \cite{BY:2003,Bulgac:2005a,ZA-SSSK:2005lr}. The phase of the superfluid order parameter changes quite abruptly by $\pi$ across a domain wall, and \DW s appear always in pairs propagating in opposite directions.  The \DW s reflect elastically from the system boundaries. However, they eventually dissipate, in particular when colliding with one another, if they have the same phase jump.  \DW s with similar phase jumps can propagate at different speeds and when they catch one another often annihilate. \DW s with opposite phase jumps appear to collide elastically. 

%%%%%%%%%%%%%%%%%%%%%%%%%%%%%%%%%%%%%%%%%%%%%%%%%%%%%%%%% 
We thank J.A. Joseph and J.E. Thomas for sharing their experimental data \cite{Joseph:2011} and  M.M. Forbes for discussions. This work was supported by DOE Grants DE-FG02-97ER41014, DE-FC02-07ER41457, DE-AC05-760RL01830. Calculations have been performed on UW Hyak (NSF MRI PHY-0922770), Franklin (Cray XT4,  NERSC, DOE B-AC02-05CH11231), and JaguarPF (Cray XT5, NCCS, DOE DE-AC05-00OR22725). 

%%%%%%%%%%%%%%%%%%%%%%%%%%%%%%%%%%%%%%%%%%%%%%%%%%%%%%%%% 

%\bibliographystyle{apsrev4-1}
%\bibliography{master}


 \end{document}